\documentclass[runningheads,a4paper]{llncs}
\usepackage{graphicx}
\usepackage[svgnames,table]{xcolor}
  \definecolor{lightgray}{gray}{0.9}
\usepackage{url}
\usepackage[colorlinks,allcolors=Blue]{hyperref} %
\usepackage{algorithm}
\usepackage{algpseudocode}
\usepackage{cleveref} 
  \Crefname{section}{Section}{Sections}
  \Crefname{table}{\tablename}{Tables}
  \Crefname{figure}{\figurename}{Figures}
  
\usepackage{booktabs}
\usepackage{multirow}
\usepackage{tabularx}
\usepackage{wrapfig}
\usepackage[english]{babel}
\usepackage{enumerate}
\usepackage{ragged2e}  %
  \newcolumntype{P}[1]{>{\raggedright\arraybackslash}p{#1}}
  \newcolumntype{?}{!{\vrule width 1pt}}
\usepackage{amsmath,amssymb}%
\usepackage{mathabx}
\usepackage{colortbl}
\usepackage{paralist}
\usepackage[font={small}]
  {caption,subfig}
\usepackage{tikz}
\usetikzlibrary{arrows,arrows.meta,automata,shapes,decorations.pathreplacing,calc}
  \usetikzlibrary{decorations.pathmorphing}
  \usetikzlibrary{angles,quotes}
  \usetikzlibrary{positioning}
\usepackage{tikzscale}
  \pgfmathdeclarefunction{gauss}{2}{%
    \pgfmathparse{1/(#2*sqrt(2*pi))*exp(-((x-#1)^2)/(2*#2^2))}%
  }
\usepackage{pgfplots}

\newcommand{\wrts}{w.r.t.\ }
\newcommand{\cf}{cf.\ } 		
\newcommand{\ies}{i.e.,\ }

\newcommand{\egs}{e.g.\ } 		
\newcommand{\Eg}{For example,\ }
\newcommand{\eg}{for example,\ }

\usepackage[colorinlistoftodos]{todonotes} %
\presetkeys{todonotes}{inline}{}

\usepackage{ctable}

\usepackage{times}

\tikzset{x=.35cm,y=.35cm}
\newcommand{\st}[2][]{{#1}_{\mathit{#2}}}
\newcommand{\stsaf}[1][]{\mathit{saf}_{\mathit{#1}}}
\newcommand{\sthaz}[1][]{\mathit{haz}_{\mathit{#1}}}
\newcommand{\stmis}[1][]{\mathit{mis}}
\newcommand{\stc}[2][]{\overline{\mathit{#2}}_{#1}}

\newcommand{\stmh}[2][]{\underline{\mathit{#2}}_{#1}}

\newcommand{\ac}[2][]{\mathit{#2}_{#1}}
\newcommand{\acendg}[2][]{\ac[#1]{e}^{\mathit{#2}}}
\newcommand{\accomp}[2][]{\ac[#1]{m}^{\mathit{#2}}}
\newcommand{\acf}[1][]{\ac{f^{#1}}}

\newcommand{\tpar}{\parallel_t}

\newcommand{\os}[1][]{\mathit{os}_{\mathit{#1}}}

\newcommand{\accompver}[1]{\mathsf{rv}_m(#1)}
\newcommand{\acendgver}[1]{\mathsf{rv}_e(#1)}
\newcommand{\CtrLoop}{\mathcal{L}}
\newcommand{\Haz}{\mathcal{H}}

\newcommand{\Os}{\mathcal{O}}
\newcommand{\Comp}{\mathcal{M}}
\newcommand{\Endg}{\mathcal{E}}
\newcommand{\RS}{\mathfrak{R}}
\newcommand{\Act}{\mathcal{A}}
\newcommand{\Weights}{\mathcal{W}}

\newcommand{\sev}[1]{\mathsf{sv}(#1)}
\newcommand{\prob}[1]{\mathsf{pr}(#1)}
\newcommand{\cost}[1]{\mathsf{cs}(#1)}
\newcommand{\riskprio}[1]{\mathsf{rp}(#1)}
\newcommand{\reach}[1]{\mathsf{reach}_{\Delta}(#1)}
\newcommand{\wsev}{\mathsf{sv}}
\newcommand{\wcost}{\mathsf{cs}}

\newcommand{\wriskprio}{\mathsf{rp}}

\newcommand{\abs}[1]{\lvert{#1}\rvert}
\newcommand{\ocomp}{\preceq_m}

\newcommand{\pstates}{\Sigma^p}
\newcommand{\sstates}{\Sigma^s}

\newcommand{\procdet}{\textsf{estimate}_{\CtrLoop,\os}}

\newcommand{\keywords}[1]{\par\addvspace\baselineskip
\noindent\keywordname\enspace\ignorespaces#1}

\urldef{\mailsa}\path|{mario.gleirscher, stefan.kugele}@tum.de|    

\begin{document} 
\mainmatter 
\title{From Hazard Analysis to Hazard Mitigation Planning:
  The Automated Driving Case%
  \thanks{The final publication is available at Springer via
    \url{http://doi.org/10.1007/978-3-319-57288-8_23}}
  }
\titlerunning{Hazard Mitigation Planning: Automated Driving}
\author{
Mario Gleirscher (\url{http://orcid.org/0000-0002-9445-6863})
\and
Stefan Kugele
}
\authorrunning{M.\ Gleirscher and S.\ Kugele}

\institute{Technische Universit\"at M\"unchen, Munich,
    Germany
\mailsa}

\toctitle{Lecture Notes in Computer Science}
\tocauthor{Authors' Instructions}
\maketitle

\begin{abstract}
  Vehicle safety depends on (a) the range of identified hazards and (b) the operational situations for which mitigations of these hazards are acceptably decreasing risk.  Moreover, with an increasing degree of autonomy, risk ownership is likely to increase for vendors towards regulatory certification.  Hence, highly automated vehicles have to be equipped with verified controllers capable of reliably identifying and mitigating hazards in all possible operational situations.  To this end, available methods for the design and verification of automated vehicle controllers have to be supported by models for hazard analysis and mitigation.

  In this paper, we describe 
  \begin{inparaenum}
  \item[(1)] a framework for the analysis and design of \emph{planners} (\ies high-level controllers) capable of run-time hazard identification and mitigation,
  \item[(2)] an incremental algorithm for constructing \emph{planning models} from hazard analysis, and
  \item[(3)] an exemplary application to the design of a \emph{fail-operational controller} based on a given control system architecture.
  \end{inparaenum}
  Our approach equips the safety engineer with concepts and steps to (2a) elaborate scenarios of endangerment and (2b) design operational strategies for mitigating such scenarios.

\end{abstract}

\section{Challenges, Background, and Contribution}
\label{sec:backgr-motiv}
\label{sec:challenges}

Automated and autonomous vehicles (AV) are responsible for avoiding mishaps and even for mitigating hazardous situations in as many \emph{operational situations} as possible.  
Hence, AVs are examples of systems where the identification (2a) and mitigation (2b) of hazards have to be highly automated.  This circumstance makes these systems even more complex and difficult to design.  Thus, safety engineers require specific models and methods for risk analysis and mitigation. 

As an example, we consider \emph{manned road vehicles in road traffic with an autopilot (AP) feature}. Such vehicles are able to automatically conduct a ride only given some valid target and minimizing human intervention.  
The following \emph{AV-level (S)afety (G)oal} specifies the problem we want to focus on in this paper:
\begin{quote}
  \textbf{SG:} The AV can always reach a \emph{safest possible state} $\sigma$ wrt.\ the hazards identified and present in a specific operational situation $\os$.
\end{quote} 

\subsubsection{Background.}
\label{sec:background}

Adopted from \cite{Ericson2015,Leveson2012}, we give a brief overview of terms used in this paper: We perceive a \emph{mishap} as an event of harm, injury, damage, or loss.  A \emph{hazard} (or \emph{hazardous state}) is an event that can lead to a mishap. We consider hazards to be factorable.  Hence, a hazard can play the role of a \emph{causal factor} of another hazard or a mishap.  We denote causal factors, hazards, and mishaps---\ies the elements of a \emph{causal (event) chain}---by the term \emph{safety risk} (\emph{risk state} or \emph{risk} for short).  We perceive the part of a causal chain increasing risk as an \emph{endangerment scenario}, and the part of a causal chain decreasing risk as a \emph{mitigation strategy}. 
\Cref{tab:hazcausalfactorcomb} exemplifies different \emph{endangerment} scenarios and how these can be \emph{mitigated} using corresponding strategies.

Mitigation strategies can be seen as specific \emph{system-level safety requirements} implemented by a given control system architecture.  We assume that a control system architecture consists of \emph{features} deployed on \emph{sensors}, \emph{actuators}, and \emph{software components} running on \emph{networked computing units} (cf.~\Cref{fig:bna}).  By \emph{traditional driver assistance (TDA)}, we refer to driver assistance features already in the field, \egs \emph{adaptive cruise control (ACC)} and \emph{lane keeping assistance (LKA)}.

We distinguish between the domains \emph{vehicle}, \emph{driver}, and \emph{road environment}.  For highly and fully automated driving, not all domains have to be considered.  For example, in full automation (\egs level 5 in \cite{sae:j3016}), the vehicle has to operate under all road and environmental conditions manageable by a human driver and therefore a driver does not have to be taken into account.

\begin{table}[t]
  \centering
  \caption{Examples of endangerment scenarios and mitigation strategies.}
  \label{tab:hazcausalfactorcomb}
  \resizebox{.9\textwidth}{!}{
\begin{tabularx}{1.0\linewidth}{lXXXX}%
  \toprule\noalign{\smallskip}
  & & \multicolumn{3}{c}{
    \textbf{Possible Mitigation Strategy}} 
  \\[.3em]
  \multicolumn{2}{c}{\multirow{2}{10em}[1.2em]{
    \textbf{Scenario of Endangerment}}} 
  & \textbf{Vehicle} 
  & \textbf{Driver} 
  & \textbf{RoadEnv} 
  \\\midrule\noalign{\smallskip}
  Vehicle  
    & subsystem\newline fault 
    & dependability\newline pattern 
    & controlled\newline shutdown 
    & car2x com.,\newline digital road signs
  \\\midrule
  Driver   
    & maloperation    
    & passive safety        
    & \multicolumn{2}{c}{safe reaction (if controllable)} 
  \\\midrule
  \multirow{2}{4em}[-1em]{RoadEnv}  
    & unforeseen\newline obstacle 
    & emergency\newline braking assistant 
    & braking or\newline circumvention
    & digital road signs, x2car com. 
  \\\cmidrule{2-5}
    & IT attack 
    & security pattern 
    & \multicolumn{2}{c}{safe reaction (if controllable)} 
  \\\bottomrule
\end{tabularx}
   }
\end{table}

\subsubsection{Contribution.}
\label{sec:contribution}

Elaborating on previous work in \cite{Gleirscher2016b,Gleirscher2017a}, we contribute
\begin{itemize}
\item[(1)] a framework for modeling, analysis, and design of \emph{planners} (\ies high-level controllers) capable of run-time hazard identification and mitigation, and
\item[(2)] a procedure for constructing \emph{planning models} from hazard analysis.
\end{itemize}
For this, we formalize the core engineering steps necessary for (2a) the identification and analysis of scenarios of endangerment and (2b) the design of operational mitigation strategies.  Using an exemplary AV, we incrementally build up a risk structure involving three hazards in the vehicle domain, 
as well as several strategies to reach safe states in presence of these hazards.  We discuss approaches to model reduction suited for run-time hazard analysis and mitigation planning where efficient identification of operational situations and acting therein play a crucial role.  

In this paper, we discuss related work in~\Cref{sec:related-work}, our abstraction in \Cref{sec:modelinggeneral}, and our modeling framework in \Cref{sec:concepts-ocp}.  \Cref{sec:procedure} shows a procedure for building a hazard mitigation planning model.  We present an AV example in~\Cref{sec:example-bmw}, discuss our approach in~\Cref{sec:discussion}, and conclude in \Cref{sec:conclusion}.

\section{Related Work}
\label{sec:related-work}

Among the related formal methods available in robotics planning, embedded systems, and automated vehicle control, we only discuss a few more recent ones and highlight how we can improve over them.

G\"udemann and Ortmeier~\cite{Guedemann2010} present a language for probabilistic system modeling for safety analysis.  Formalized as \textsc{Markov} \emph{decision processes (MDP)}, they propose two ways of failure mode modeling (\ies per-time and per-demand failure modes), and two ways of deductive cause consequence reasoning (\ies quantitative and qualitative).  Their model and reasoning can extend our approach.  However, our work (i) adds stronger guidelines on how to build planning models and (ii) puts hazard analysis into the context of autonomous systems and mitigation planning.

Eastwood et al.~\cite{Eastwood2016} present an algorithm for finding permissive robot action plans optimal \wrts to safety and performance.
They employ \emph{partially observable MDPs} (helpful in regarding uncertainty and robot limitations) to model robot behavior, and two abstractions from this model to capture a system's modes and hazards.
Our framework uses three layers of abstraction ($\Sigma^s$, $\Sigma^p$, $\Sigma$), operational situations to capture control modes, and a structure to capture hazards. 
While they directly encode hazard severity for plan selection, our framework allows the planner to calculate the risk priority based on a causal event tree towards mishaps.
As opposed to complete behavioral planning, our approach focuses the construction of mitigation planning models.  For example, for system faults we can plan mitigations by using adaptation mechanisms of a given control system architecture.

Jha and Raman~\cite{Jha2016} discuss the synthesis of vehicle trajectories from probabilistic temporal logic assertions.  Synthesized trajectories take into account perception uncertainty through approximation of sensed obstacles by combining Gaussian polytopes. 
In a similar context, Rizaldi and Althoff~\cite{Rizaldi2015} formalize safe driving policies to derive safe control strategies implementing worst-case braking scenarios in autonomous driving.  They apply a hybrid-trace-based formalization of physics required for model checking of recorded \cite{Rizaldi2015} and planned \cite{Rizaldi2016} strategies.  
\cite{Jha2016,Rizaldi2015,Rizaldi2016} discuss low-level control for a specific class of driving scenarios, whereas our approach provides for 
(i) the investigation and combination of many related operational situations, thus, forming a more comprehensive perspective of driving safety,
(ii) regarding various kinds of hazards that might play a role in high- and low-level control beyond safe and optimal trajectory planning and collision avoidance.

Wei~et~al.~\cite{Wei2013} describe an autonomous driving platform, capable of bringing vehicles to a safe state and stop, \ies activating a fail-operational mode on critical failure, and a limp-home mode on less critical failure.  These are mitigation strategies we can assess in our framework.  Their work elaborates on designing a specific class of architectures.  Additionally, we provide an approach to systematically evaluate risks and, consequently, derive an architecture design.

Babin~et~al.~\cite{Babin2016} propose a system reconfiguration approach developed with the Event-B method in a correct-by-construction fashion using a behavior pattern similar to our approach (particularly, \Cref{fig:statemodel}).  Reconfiguration as one way to \emph{mitigate} faults is discussed in this work.  
Wardzi\'{n}ski~\cite{Wardzinski2008} discusses hazard identification and mitigation for autonomous vehicles by predetermined risk assessment (\ies with safety barriers) and dynamic risk assessment.  For both, he provides argumentation patterns for creating AV safety cases.  In addition to his work, the abstraction and the method we propose covers both paradigms in one framework.  We provide formal notions of all core concepts.

\section{Abstraction for Run-time Hazard Mitigation}
\label{sec:modelinggeneral}
\label{sec:abstractions}

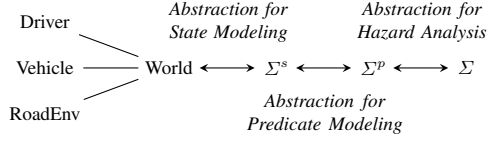
\begin{figure}[t]
  \centering
\begin{tikzpicture}[
  scale=.8, every node/.style={transform shape},
  >=stealth,node distance=1em and 3.0em, shorten >=1pt,auto]
  \node[] (worldV) {Vehicle};
  \node[right       =of worldV] (world) {World};  
  \node[above       =of worldV] (worldD)  {Driver};
  \node[below       =of worldV] (worldRE) {RoadEnv};
  \node[right       =of world]  (sigmas)  {$\sstates$};
  \node[right       =of sigmas] (pstates)  {$\pstates$};
  \node[right       =of pstates](haz)     {$\Sigma$};
  \draw (world) -- (worldD);
  \draw (world) -- (worldV);
  \draw (world) -- (worldRE);  

  \draw[<->] (world) -- node[text width=8em,above=1em,align=center]{\emph{Abstraction for\\State Modeling}} (sigmas); 
  \draw[<->] (sigmas) -- node[text width=12em,below=1em,align=center]{\emph{Abstraction for\\Predicate Modeling}} (pstates);  
  \draw[<->] (pstates) -- node[text width=10em,above=1em,align=center]{\emph{Abstraction
      for\\Hazard Analysis}} (haz); 
\end{tikzpicture}
   \caption{Abstractions for \emph{state} and \emph{predicate} modeling, and for \emph{hazard analysis}.}
  \label{fig:abstractions}
\end{figure}

\Cref{fig:abstractions} depicts three \emph{abstractions}---$\Sigma^s$, $\Sigma^p$, and $\Sigma$---for run-time hazard mitigation in AVs. 
The \emph{state space} $\Sigma^s$ pertains to the quantization of continuous signals from the physical \emph{world} encompassing the \emph{driver} ($\mathtt{drv}$), the \emph{vehicle} ($\mathtt{veh}$), and the \emph{road environment} ($\mathtt{renv}$).  For instance, the quantity \emph{speed} is represented by the discrete state variable $\mathtt{veh.speed}$, which in turn is used to formulate predicates to obtain the \emph{abstract state space} $\Sigma^p$.  For example, a predicate over sensor values $p(\mathtt{veh.speed}$, $\mathtt{veh.loc}$, $\mathtt{renv.map})$ can encode $\mathit{exitTunnel}$, an invariant constraining the activity of leaving a tunnel. 
We describe this two-staged abstraction in more detail in \cite{Gleirscher2017a}.  

Here, we will work with the \emph{risk state space} $\Sigma$ whose concepts---actions, hazard phases, their composition and ordering---are discussed below:

\subsubsection{Actions.}
\label{sec:actions}

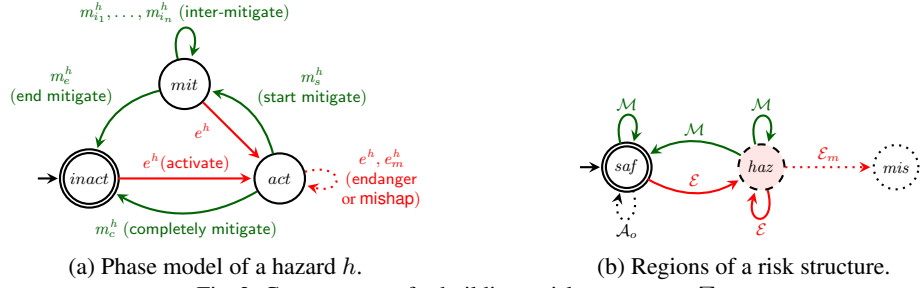
\begin{figure}[t]
  \subfloat[Phase model of a hazard $h$.]{
\begin{tikzpicture}[
  ->,>=stealth,thick,node distance=8em,
  scale=.7,every node/.style={transform shape},
  shorten >=1pt,auto,initial text={},
  every state/.style={minimum width=3em}]
  \node[state,initial,accepting] (0) {$inact$}; 
  \node[state] (c) [above right of=0] {$mit$}; 
  \node[state] (1) [below right of=c] {$act$}; 
  \path 
  (0) edge[red]  node[align=center] {$\acendg{h} (\mathsf{activate})$} (1) 
  (1) edge[bend right,swap,DarkGreen] node[align=center] {$\accomp[s]{h}$\\$(\mathsf{start\;mitigate})$} (c)
  (c) edge[bend right,DarkGreen] node[swap,align=center] {$\accomp[e]{h}$\\$(\mathsf{end\;mitigate})$} (0)
  (1) edge[bend left,DarkGreen] node[align=center] {$\accomp[c]{h}$
    $(\mathsf{completely\;mitigate})$} (0)
  (c) edge[red] node[near start,swap,align=center] {$\acendg{h}$} (1)
  (1) edge[loop right,red,dotted] node[align=center] {$\acendg{h},\acendg[m]{h}$\\$(\mathsf{endanger}$\\or $\textsf{mishap})$} (1)
  (c) edge[loop above,DarkGreen] node[align=center] {$\accomp[i_1]{h},\ldots,\accomp[i_n]{h}$ $(\mathsf{inter\mbox{-}mitigate})$} (c);
\end{tikzpicture}
     \label{fig:hazardphases}}
  \hfill
  \subfloat[Regions of a risk structure.]{
\begin{tikzpicture}[
  ->,>=stealth,thick,node distance=8em,
  scale=.7,every node/.style={transform shape},
  shorten >=1pt,auto,initial text={}]
  \node[state,initial,accepting] (saf) {$\stsaf$}; 
  \node[state,fill=red!10,dashed] (haz) [right of=saf] {$\sthaz$}; 
  \node[state,dotted] (mis) [right of=haz] {$\stmis$}; 
  \path 
  (saf) edge [dotted,loop below] node {$\Act_o$} (saf) %
  (saf) edge [bend right,red]  node {$\Endg$}      (haz) 
  (haz) edge [bend right,swap,DarkGreen] node {$\Comp$}
  (saf)
  (haz) edge[loop below,red] node {$\Endg$} (haz)
  (haz) edge[loop above,DarkGreen] node {$\Comp$} (haz)
  (saf) edge[loop above,DarkGreen] node {$\Comp$} (saf)
  (haz) edge[dotted,red]          node {$\Endg_m$} (mis);
\end{tikzpicture}
     \label{fig:statemodel}}
  \caption{Core concepts for building a risk state space $\Sigma$.
    \label{fig:riskstatespace}}
\end{figure}

Let $\Act$ be a set of actions. We abstract from \emph{control loop behaviors} within and across operational situations by distinguishing four classes of actions:  \emph{endangerments} $\Endg$, \emph{mitigations} $\Comp$~(see \Cref{fig:statemodel}), \emph{mishaps} $\Endg_m$, and \emph{ordinary actions} $\Act_o$.  Note that actions can take place in one or more out of the three domains, \texttt{drv}, \texttt{veh}, and \texttt{renv}, depending on the quantities they modify.  We require $\Endg,\Comp,\Act_o,\Endg_m\subset\Act$.

\begin{definition}[Hazard Phases]
  \label{def:hazard-phases}
  Let $\Haz$ be a set of hazards.
  Given $h\in\Haz$,
  endangerment actions $\acendg{h},\acendg[m]{h}\in\Act$, and
  $n_h\in\mathbb{N}\setminus\{0\}$ mitigation actions $\accomp[j]{h}\in\Act$,
  we define the \emph{phases of a hazard} $h$ 
  as the set 
  $P_h = \{0,\acendg{h},\acendg[m]{h}\} \cup 
         \{m^h_j\mid j\in\mathbb{N}\setminus\{0\}\land j\leq n_h\}$ whose elements denote the following:
  \begin{itemize}
  \item[$0$] \hspace{1em} hazard $h$ is (inact)ive,
  \item[$\acendg{h}$] \hspace{1em} hazard $h$ has been (act)ivated by an action $\acendg{h}$, 
  \item[{$\acendg[m]{h}$}] \hspace{1em} (act)tivated hazard $h$ has contributed to a mishap by an action $\acendg[m]{h}$, and
  \item[{$\accomp[j]{h}$}] \hspace{1em} hazard $h$ has been (mit)igated by an action $\accomp[j]{h}$.
  \end{itemize}
\end{definition}
For each hazard $h$, \Cref{fig:hazardphases} depicts $P_h$ as a transition system where $\abs{P_h}=n_h+3$, the indices $s,e,c,i_1,\ldots,i_n\leq n_h$, the state $mit$ subsumes $n_h-1$ phases, $act$ subsumes phases $\st[e]{}^h$ and $\st[e]{m}^h$.
\Eg in the vehicle domain, $\accomp[s]{h}$ can model \emph{degradation} transitions and $\accomp[e]{h}$ or $\accomp[c]{h}$ can model \emph{repair} transitions.

From all the sets of hazard phases, we compose a tuple space as follows: 

\label{sec:states}
\begin{definition}[Risk State Space]
  \label{def:riskstates}
  Based on \Cref{def:hazard-phases}, we define the \emph{risk state space} $\Sigma$ as the set of $|\Haz|$-tuples
  \[ \{ (p_{h_1}, \ldots,p_{h_{|\Haz|}}) \mid \forall i \in \{1,
  \ldots, |\Haz|\}: h_i \in\Haz \land p_{h_i} \in P_{h_i} \}\enspace. \]
\end{definition}

We call any subset of $\Sigma$ a \emph{region}. Let $\sigma, \sigma' \in\Sigma$ with $\sigma=(p_{h_1},\ldots,p_{h_{\abs{\Haz}}})$ and $\sigma'=(p'_{h_1},\ldots,p'_{h_{\abs{\Haz}}})$.  
To quantify risk in \emph{scenarios of endangerment} and \emph{mitigation strategies}~(\Cref{tab:hazcausalfactorcomb}), we define a partial order over $\Sigma$:

\begin{definition}[Mitigation Order] 
  \label{def:comp_preorder}
  Let $P_h$ be a set of phases for hazard $h$ (\Cref{def:hazard-phases}) and 
  $\prec_h\; = \{(e^h,0),(e^h,m^h_j),(m^h_j,0),(e^h_m,e^h)
  \mid m^h_j\in P_h \}$.  
  By the reflexive transitive closure\footnote{Here, for a relation $R$, $R^n$ represents the composition of relations.} $\preceq_h = \{(p,p)\mid p\in P_h\} \cup \bigcup_{n\geq 1} \prec_h^n$, 
we define the \emph{mitigation order} $\ocomp\;\subseteq\Sigma\times\Sigma$,
  for \emph{states}
  $\sigma, \sigma' \in\Sigma$, as follows:
  \[ \sigma \ocomp \sigma' \Leftrightarrow \forall i\in \{1,\ldots,|\Haz|\}: p_{h_i} \preceq_h p'_{h_i} \enspace. \]
\end{definition}
Intuitively, $\sigma\prec_m\sigma'$ denotes ``$\sigma'$ is better or further in mitigation than $\sigma$.''\footnote{We use the convention $\sigma\prec_m\sigma'\equiv\sigma\ocomp\sigma'\land\sigma\neq\sigma'$.}  

\section{Concepts for Run-time Hazard Mitigation}
\label{sec:concepts-ocp}

In this section, we explain the core concepts of deriving a \emph{risk structure} for a specific \emph{operational situation}.  Using the risk state space $\Sigma$ and actions $\Act$, we define the notions of \emph{risk structure}, \emph{risk region}, and \emph{operational situation}:

\begin{definition}[Risk Structure]
  \label{def:riskstructure}
  A \emph{risk structure} is a weighted labeled transition system $(\Sigma,\Act,\Delta,\Weights)$ with
  \begin{itemize}
  \item a set $\Sigma$ called the \emph{risk state space} (\Cref{def:riskstates}),
  \item a set $\Act$ of \emph{actions} used as transition labels, 
  \item a relation $\Delta\subseteq\Sigma\times\Act\times\Sigma$ called \emph{labeled transition relation}, and
  \item a set $\Weights$ of partial functions $w:(\Sigma\cup\Act\cup\Delta)\rightarrow\mathbb{W}_w$ called \emph{weights} where the set $\mathbb{W}_w$ can be, \egs $\mathbb{N, R},[0,1]$, or $\{m,c,f\}$.%
    \footnote{(m)arginal, (c)ritical, (f)atal; for other examples of severity scales, see \cite{Ericson2015}.}
  \end{itemize}
\end{definition}
To capture the notions of endangerment scenario and mitigation strategy (\Cref{tab:hazcausalfactorcomb}) based on $\Delta$, we consider paths and strategies:

\begin{definition}[Paths, Strategies, and Reachability]
  \label{def:strategy}
  By convention, we write $\sigma\stackrel{a}{\longrightarrow}\sigma'$ for $(\sigma,a,\sigma')\in\Delta$.  Then, for $n,l\in\mathbb{N}\setminus\{0\}$, a \emph{path} is a sequence $\sigma_0\stackrel{a_0}{\longrightarrow}\ldots\sigma_{n-1}\stackrel{a_{n-1}}{\longrightarrow}\sigma_n$.  
  By $\Delta^l$ %
  we denote the set of all paths of length $l$ and by $\Delta^{\infty}=\bigcup_{l>0}\Delta^l$ all paths over $\Delta$.
  Furthermore, we call a set $S\subset\Delta^{\infty}$ a \emph{strategy}.
By $\mathsf{reach}_{\Delta}:\Sigma\rightarrow 2^{\Sigma}$ with 
$\mathsf{reach}_{\Delta}(\sigma) = 
  \{\sigma\} \cup 
  \{\sigma'\in\Sigma\mid \exists 
  \sigma\stackrel{a}{\longrightarrow}\ldots 
  \stackrel{a'}{\longrightarrow}\sigma' 
  \in\Delta^{\infty}\}$, 
we denote the set of states reachable in $\Delta$ from a state $\sigma$.
\end{definition}

\paragraph{Endangerments.}
\label{sec:hazard}

We consider an action $a\in\Act$ as an \emph{endangerment}, \ies $a\in\Endg$, if $\sigma\succ_m\sigma'$ for a transition $(\sigma,\ac{a},\sigma')\in\Delta$.  The class $\Endg$ models steps of endangerment scenarios.  \Eg $a$ can stem from faults in \texttt{drv}, \texttt{veh}, and \texttt{renv}.

\paragraph{Mitigations.}
\label{sec:reaching-safe-state}

We consider an action $a\in\Act$ as a \emph{mitigation}, \ies $a\in\Comp$, if $\sigma\prec_m\sigma'$ for a transition $(\sigma,\ac{a},\sigma')\in\Delta$.  The class $\Comp$ models steps of mitigation strategies.  One objective of a good mitigation strategy is to achieve a \emph{stable safe state}.  

\paragraph{Operational Situations.}
\label{sec:os}
States and regions in $\Sigma$ both correspond to subsets of $\Sigma^s$~(\Cref{sec:modelinggeneral}).  To limit the scope of a risk analysis, we use an \emph{operational situation} which combines an \emph{initial region} with a \emph{(reasonably weak) invariant} holding along the driving scenarios in a specific road environment.

\begin{definition}[Operational Situation]
  \label{def:os}
  An \emph{operational situation} is a tuple $(\Sigma_0,\{ \sigma\in\Sigma^s \mid p(\sigma) \})$ where $\Sigma_0\subseteq\Sigma$ and $p$ is an invariant over $\Sigma^s$ including all representations of $\Sigma_0$ in $\Sigma^s$.  Let $\Os$ be the set of all operational situations.
\end{definition}
Below, we will work with a risk structure $\RS_{\os}=(\Sigma,\Act,\Delta,\Weights)$ and assume a fixed operational situation $\os\in\Os$ associated with $\RS_{\os}$. Hence, we use $\RS$ solely. 

\subsubsection{Risk Regions.}
\label{sec:regions}

We consider specific subsets of $\Sigma$ called \emph{risk regions}, particularly, the \emph{safe region} $\stsaf$, the \emph{hazardous region} $\sthaz$, and the \emph{mishap region} $\stmis$~(see \Cref{fig:statemodel}).  Safety engineers aim at the design of mitigations which (i) avoid $\stmis$ and (ii) react to endangerments as early and effectively as possible.  Then, $\Endg_m$ reduces to unavoidable actions from so-called \emph{near-mishaps} still in $\sthaz$ towards $\stmis$.  \Eg we consider a successfully deployed airbag to be in $\Comp$ such that $\stmis$ is not reached in such an accident~(more in \Cref{sec:discussion}).  

Our definitions of risk regions depend on $\RS$:  First, $\stmis = \{ (p_{h_1},\ldots,p_{h_{\abs{\Haz}}}) \in \Sigma \mid \exists i \in \{1,\ldots,|\Haz|\}\colon p_{h_i}=e^{h_i}_m \}$.  We require mishaps to be \emph{final}, \ies $\forall\sigma\in\stmis\colon\mathsf{reach}_{\Delta}(\sigma) = \{\sigma\}$.  Second, $\stsaf$ and $\sthaz$ vary with a given operational situation.  Moreover, they can be defined based on, \egs weights and equivalences.  However, $(0,\ldots,0)\in\stsaf$ and, for an $\os$, we start in the safe region iff $\Sigma_0\subseteq\stsaf$.

\paragraph{Weights.}
\label{sec:weights}

By associating \emph{weights} with elements of $\RS$, we quantify further details on the physical phenomena of the controlled process relevant for risk analysis.  

\Eg given $\delta=(\sigma,\acendg{h},\sigma')\in\Delta$ with $\acendg{h}\in\Endg$, the \emph{probability of endangerment} $\prob{\delta}\in[0,1]$ yields the probability that hazard $h$ gets activated in $\sigma'$ by performing $\acendg{h}$ in  $\sigma$.
Furthermore, given $\delta=(\sigma,\accomp[j]{h},\sigma')\in\Delta$ with $\accomp[j]{h}\in\Comp$,
\begin{itemize}
\item the \emph{probability of mitigation} $\prob{\delta}\in[0,1]$ yields the probability that hazard $h$ gets mitigated in $\sigma'$ by performing $\accomp[j]{h}$ in $\sigma$.
\item the \emph{cost of mitigation} $\cost{\delta}\in\mathbb{N}$ yields the potential effort (\ies time, energy, other resources) of performing the mitigation $\accomp[j]{h}$.
\end{itemize}
For any mishap $\sigma\in\stmis$,  $\mathsf{sv}(\sigma)\in\{m,c,f\}$ specifies its \emph{severity}.  Depending on the abstraction, we can use qualitative (as shown above) or quantitative scales for $\wsev$ and $\wcost$.  Anyway, we assume to have operators for $\wsev$ and $\wcost$, \egs see \Cref{tab:svoperators}.

Weights are typically calculated from measurements of the controlled process. For example, the estimation of $\prob{\sigma,\accomp[j]{h},\sigma'}$ might be result of a \emph{controllability analysis} of $\accomp[j]{h}$ in $\sigma$ (of an operational situation).
Moreover, further quantities (\egs risk priority) might be (i) calculated from weights, (ii) be propagated along $\Delta$, and (iii) lead to an update of weights.

\subsubsection{Risk Priority.}
\label{sec:riskprio}

Given $\sigma\in\Sigma, \stmis'\subseteq\stmis$, and a function $\wriskprio :\Sigma\rightarrow\{m,c,f\}$, we can compute the  \emph{minimum partial risk priority}
\begin{equation}
  \riskprio{\sigma} = 
  \mathsf{Pr}(\sigma\rightarrow\Diamond\stmis')
  \cdot_{\wsev}
  \mathsf{min}\{\sigma'\in  
    (\stmis'\cap\mathsf{reach}_{\Delta}(\sigma))
  \mid\sev{\sigma'}\}
  \label{eq:riskprio}
\end{equation}

where $\mathsf{Pr}(\sigma\rightarrow\Diamond\stmis')\in[0,1]$ denotes the probability\footnote{See, \egs \cite{Baier2008} for details about probabilistic temporal logic and reasoning.} that from $\sigma$ some mishap $\sigma'\in\stmis'$ is eventually ($\Diamond$) reached in $\RS$. This definition implements a traditional measure of risk analysis (see, \egs \cite{Ericson2015}), referring to the \emph{minimum negative outcome} (\ies damage, injury, harm, loss) possibly reachable from $\sigma$ in a specific operational situation $os\in\Os$.
Note that for $\sigma\in\stmis$, $\riskprio{\sigma}=\sev{\sigma}$.

\subsubsection{Equivalences over $\Sigma$.}
For simplification of complex risk structures $\RS$, we can construct equivalence classes over states.  From the structure of states in $\Sigma^s$, the dynamics in $\Sigma^s$, and the elements of the control system architecture (\Cref{sec:background}), we give a brief informal overview of equivalences over $\Sigma$ to be considered:

We speak of \emph{feature equivalence}, $\sigma\approx_f\sigma'$, iff both, $\sigma$ and $\sigma'$ map to the same set of \emph{active features} of the control system, %
\ies \emph{in-the-loop} no matter whether they are fully operational, faulty, or degraded. Note that out-of-the-loop features can be faulty, deactivated, or in standby mode.
Next, we speak of \emph{degradation equivalence}, $\sigma\approx_d\sigma'$, iff $\sigma\approx_f\sigma'$ and both states share the same set of \emph{degraded features}.
Furthermore, we speak of \emph{hazard (or fault) equivalence}, $\sigma\approx_h\sigma'$, iff $\forall i \in \{1,\ldots,|\Haz|\}: p_{h_i} \in P_{h_i}\setminus\{0\} \Leftrightarrow p_{h_i}' \in P_{h_i}\setminus \{0\}$, 
and, particularly, of \emph{mishap equivalence}, $\sigma\approx_{h_m}\sigma'$, iff
$\forall i \in \{1,\ldots,|\Haz|\}: p_{h_i} = e^{h_i}_m \Leftrightarrow p_{h_i}' = e^{h_i}_m$.
Based on $\approx_h$, we finally define:

\begin{definition}[Mitigation Equivalence]
  \label{def:comp-equiv}
  Based on \Cref{def:comp_preorder}, two states $\sigma, \sigma'\in\Sigma$ are
  \emph{mitigation equivalent}, written $\sigma\approx_m\sigma'$, iff
  \[\sigma\approx_h\sigma'
  \land \forall i\in \{1,\ldots,\abs{\Haz}\}: p_{h_i}\succ_h e^{h_i} \Leftrightarrow
  p'_{h_i}\succ_h e^{h_i}\enspace. \]
\end{definition}

\section{Construction of Risk Structures}
\label{sec:procedure}

\begin{figure}[t]
\subfloat[The operators $=_{\wsev}, <_{\wsev}, >_{\wsev}$ and $\cdot_{\wsev}$ where $\geq_{\wsev} \;\equiv\; =_{\wsev} \land >_{\wsev}$.]{
  \label{tab:svoperators}

  \vspace{.4em}

  \begin{tabular}[b]{cccc}
    \toprule
      & m & c & f \\\midrule
    m & $=_{\wsev}$ & $<_{\wsev}$ & $<_{\wsev}$ \\
    c & $>_{\wsev}$ & $=_{\wsev}$ & $<_{\wsev}$ \\
    f & $>_{\wsev}$ & $>_{\wsev}$ & $=_{\wsev}$ \\\bottomrule
  \end{tabular}
  \hspace{.1em}
  \begin{tabularx}{1.7cm}[b]{cXXX}
    \toprule
    $\cdot_{\wsev}$ & m & c & f \\\midrule
    l & m & m & m \\
    m & m & m & c \\
    h & m & c & f \\\bottomrule
  \end{tabularx}
}
\hfill
\subfloat[Scheme for incremental construction of $\RS$ by $\mathsf{constructRS}$.]{
  \includegraphics[width=.6\linewidth]
    {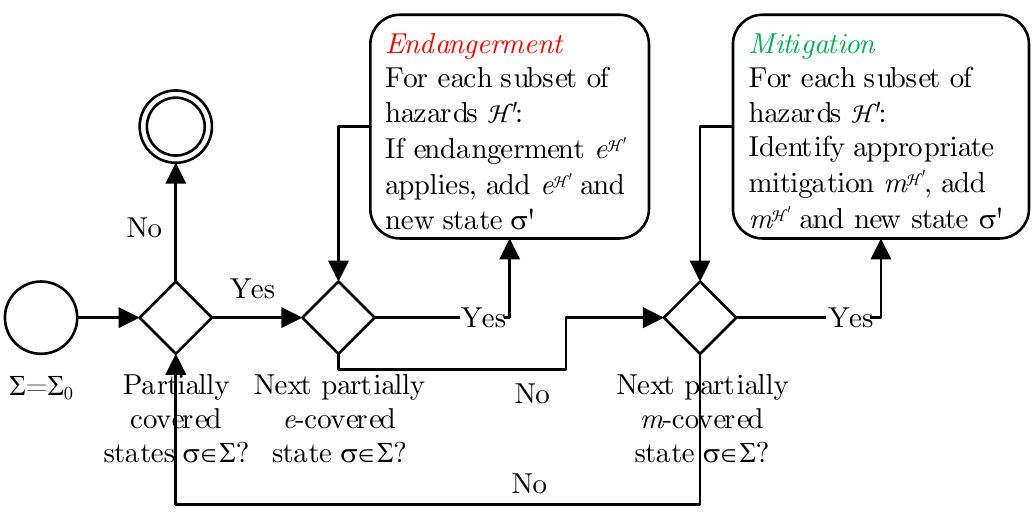}
  \label{fig:scheme-constriskgraph}
}
\caption{Operators and scheme}
\end{figure}

In this section, we describe an incremental and forward\footnote{For generation of $\RS$, backward reasoning is the alternative not shown here.} reasoning approach to building a risk structure $\RS$.  

\subsubsection{Identification of Hazards.}
\label{sec:hazid}
Throughout the construction of $\RS$, we assume to have a procedure $\mathsf{hazId}$ for the identification of a set of hazards $\Haz$ based on a fixed control loop design $\CtrLoop$ of a class of AVs and their environments, and a fixed set $\Os'\subset\Os$ of operational situations~(\Cref{def:os}).  Failure mode effects and fault-tree analysis (see, \egs\cite{Ericson2015}) incorporate widely practiced schemes for $\mathsf{hazId}$.

\subsubsection{Building the Risk Structure.}
\label{sec:constriskstruct}
\Cref{fig:scheme-constriskgraph} shows the main steps of a procedure $\mathsf{constructRS}$ which,  given a set $\Haz$ and after termination, returns all elements of a \emph{complete} risk structure $\RS$. Here, completeness is relative to $\Haz$ and means that $\RS$ can no more be extended by
\begin{inparaenum}[(i)]
\item states which are reachable by existing actions in $\Act$, 
\item actions which allow reaching non-visited states in $\Sigma$, 
\item transitions in $\Delta$ which are technically possible and probable, and
\item further knowledge by extending the domains of weights.
\end{inparaenum}
Based on \Cref{fig:scheme-constriskgraph}, \Cref{alg:consriskgraph} refines $\mathsf{constructRS}$ for a  control loop $\CtrLoop$ and an operational situation $\os\in\Os'$.

\begin{algorithm}[h]
  \small
  \caption{$\mathsf{constructRS}(\CtrLoop, \os)$} %
  \label{alg:consriskgraph}
  \begin{algorithmic}[1]
    \State $\Sigma=\Sigma_0, 
      \forall\sigma\in\Sigma_0\colon 
        \accompver{\sigma}=\acendgver{\sigma}=\emptyset$
      \While{$\Haz=\mathsf{hazId(\CtrLoop,\os)}$ and
        $\exists\sigma\in\Sigma\setminus\stmis:
        \Haz\setminus(\acendgver{\sigma} \cup 
        \accompver{\sigma})\neq\emptyset$}
        \label{alg:while}
        \ForAll{$\sigma\in\Sigma\setminus\stmis$ and
          $\Haz'\subseteq\Haz\setminus\acendgver{\sigma}$}
          \Comment{\textbf{extend endangerments}}
          \label{alg:forendg}
          \If{$(\sigma',\acendg[j]{\Haz'}) \gets  
            \mathsf{activate}(\sigma,\Haz')$}
            \Comment{state/$j^{th}$ action estab.\ $\Haz'$ or mishap}
            \label{alg:genendg}
            \State $\delta\gets (\sigma,\acendg[j]{\Haz'},\sigma')$
            \If{$\mathsf{poss}(\delta)$}
              \Comment{\textcolor{red}{add endangerment?}}
              \label{alg:possendg}
              \State $(\Sigma,\Endg,\Delta,\acendgver{\sigma'})
                \gets(
                \Sigma\cup\{\sigma'\},
                \Endg\cup\{\acendg[j]{\Haz'}\}, 
                \Delta\cup\{\delta\},
                \emptyset)$ 
                \label{alg:addendg}
              \If{$\sigma'\in\stmis$} 
                \State $\sev{\sigma'}\gets\procdet(sv,\sigma')$ 
                \Comment{severity of mishap}
                \label{alg:sevmis}
              \EndIf
              \State $\mathsf{pr}(\delta)\gets\procdet(pr,\delta)$ 
                \Comment{probability of endangerment}
                \label{alg:probendg}
            \EndIf
          \Else{}
            \Comment{$\mathsf{activate}$ returns empty tuple}
            \State $\acendgver{\sigma}\gets 
              \acendgver{\sigma}\cup\Haz'$
              \Comment{\ies $\Haz'$ activated and mishap added}
              \label{alg:endgoptioncheck}
          \EndIf
        \EndFor
          \ForAll{$\sigma\in\Sigma\setminus\stmis$ and $\Haz'\subseteq\Haz\setminus\accompver{\sigma}$}
            \Comment{\textbf{extend mitigations}}
            \label{alg:formit}
            \If{$(\sigma',\accomp[j]{\Haz'})\gets
              \mathsf{mitigate}(\sigma,\Haz')$}
              \Comment{state/$j^{th}$ action mitig.\
                $\Haz'$ from $\sigma$} 
              \label{alg:genmit}
              \State $\delta \gets 
                (\sigma,\accomp[j]{\Haz'},\sigma')$
              \If{$\mathsf{poss}(\delta)$}
                \Comment{\textcolor{DarkGreen}{add mitigation?}}
                \label{alg:possmit}
                \State $(\Sigma,\Comp, \Delta,\accompver{\sigma'})
                  \gets(
                  \Sigma\cup\{\sigma'\},
                  \Comp\cup\{\accomp[j]{\Haz'}\}, 
                  \Delta\cup\{\delta\},
                  \emptyset)$ 
                  \label{alg:addmit}
                \State $\mathsf{pr}(\delta) \gets 
                  \procdet(pr,\delta)$ 
                  \Comment{probability of mitigation}
                  \label{alg:probmit}
                \State $\mathsf{cs}(\delta)\gets\procdet(cs,\delta)$
                  \Comment{cost of mitigation} 
                  \label{alg:costmit}
              \EndIf
            \Else{}
              \Comment{$\mathsf{mitigate}$ returns empty tuple}
              \State $\accompver{\sigma} \gets 
                  \accompver{\sigma}\cup\Haz'$
              \Comment{\ies all options for $\Haz'$ are checked}
              \label{alg:mitoptioncheck}
            \EndIf
          \EndFor
        \State $\Sigma\gets \Sigma\setminus\{\sigma\in\Sigma\mid 
          \sigma\not\in\bigcup_{\sigma_0\in\Sigma_0}\reach{\sigma_0} \}$
        \Comment{removing unreachable states} 
        \label{alg:modelreduction}
        \State \dots
        \Comment{further simplifications}
      \EndWhile
    \State\Return $(\Sigma, \Endg \cup \Comp, \Delta, \{\mathsf{sv}, \mathsf{pr}, \mathsf{cs}\})$
  \end{algorithmic}
 \end{algorithm}

The \textbf{while-loop} (\cf line \ref{alg:while}) accounts for the alternation between adding endangerments and mitigations. 
By using the maps $\mathsf{rv}_e$ and $\mathsf{rv}_m$ (\cf lines \ref{alg:while}, \ref{alg:forendg}, \ref{alg:endgoptioncheck}, \ref{alg:formit}, \ref{alg:mitoptioncheck}), the algorithm keeps track of the $e$n\-dan\-ger\-ment- and $m$itigation-coverage of visited states, \ies for which hazards $\sigma$ has already been visited. 

We assume to have
(i) a function $\procdet$ (\cf lines \ref{alg:sevmis}, \ref{alg:probendg}, \ref{alg:probmit}, \ref{alg:costmit}) which acts as an oracle for weights~(\Cref{sec:weights}) depending on $(\CtrLoop,\os)$, and
(ii) a function $\mathsf{poss}$ (\cf lines \ref{alg:possendg}, \ref{alg:possmit}) which acts as an oracle for determining the technical possibility of newly identified transitions.

The first \textbf{for-loop} checks for the addition of new transitions to $\Delta$ (\cf line \ref{alg:addendg}).  The transition constructor $\mathsf{activate}$ returns a state with the given hazard or mishap activated (\ies phases $e^h$ or $e^h_m$). Note that $\mathsf{activate}$ can generate $\sigma'\in\stmis$ reachable via $e^{\Haz'}_m\in\Endg_m$.

The second \textbf{for-loop} checks for the addition of new transitions to $\Delta$ (\cf line \ref{alg:addmit}).  The transition constructor $\mathsf{mitigate}$ returns a state with the given hazards $\Haz'$ mitigated to a new phase $m^h_{h_i} \in P_h$ for each $h\in\Haz'$.  

Note that none of the constructors is idempotent, $\mathsf{mitigate}$ can construct several mitigation phases for each hazard (\cf lines \ref{alg:genmit}, \ref{alg:mitoptioncheck}) and $\mathsf{activate}$ can construct two activation phases, $e^h$ and $e^h_m$, both with the corresponding actions (\cf lines \ref{alg:genendg}, \ref{alg:endgoptioncheck}).

\subsubsection{Model Reduction.}
To keep reasoning efficient, we have to apply reachability-preserving simplifications to $\RS$ (\cf lines \ref{alg:modelreduction}f), \egs equivalences such as in~\Cref{def:comp-equiv}.  The mitigation order~(\Cref{def:comp_preorder}) helps in reducing the state space and in merging actions modifying phases of the same hazards (\ies by hazard equivalence).

\vspace{-2em}

\subsubsection{Abstraction from Control System Architecture.} %
In both stages of \Cref{alg:consriskgraph}, we need to analyze the given or envisaged architecture and to identify state variables, \egs for software modules, at an appropriate level of granularity.  

In the endangerment stage (lines \ref{alg:forendg}ff), we can perform dependability analyses to identify events that can activate causal factors. Off-line, we then design specific measures to reach the safe region again, and, on-line, we design generic measures to be refined at run-time.

Moreover, the mitigation stage (lines \ref{alg:formit}ff) helps to revise a control system architecture, \egs by adding redundant execution units and degradation paths.  Moreover, we can pursue off-line synthesis of respective parts of the control system architecture.

\subsubsection{Hazard Mitigation Planning.}
\label{sec:planners}

First, $\mathsf{hazId}$ is hybrid in the sense that it 
(i) performs the sensing of already known endangerment scenarios (\egs near-collision detection, component fault diagnosis) \emph{on-line}, and
(ii) allows the addition of new scenarios from \emph{off-line} hazard analysis.

Second, a simple \emph{planner} would continuously perform shortest weighted path search in $\RS$ to keep a list of all available lowest-risk mitigation paths (\Cref{def:strategy}) and coordinate optimized lower-level controllers.

Based on these two steps, we assume $\RS$ to be continuously updated according to the available information (\ies adding or modifying endangerments and mitigations according to known scenarios).  It is important to have powerful and precise update mechanisms, highly responsive actuation, and short control loop delays.  Main issues of signal processing are briefly mentioned in \Cref{sec:discussion}.

The notion of \emph{safest possible state}~(\textbf{SG}, \Cref{sec:challenges}) is governed by the accuracy of $\sstates$~(\Cref{sec:abstractions}), the completeness of the results of $\mathsf{hazId}$, and the exhaustiveness of $\RS$ for a fixed setting $\CtrLoop,\os$.  According to \Cref{def:comp_preorder}, for a pair $(\sigma,\sigma')\in\Sigma\times\Sigma$, we might say that $\sigma'$ is the \emph{safest possible state} iff we have
\begin{equation}
\not\exists \sigma''\in\mathsf{reach}_{\Delta_{\Comp}}(\sigma)\colon \sigma'\prec_m\sigma''
\label{eq:safpassstat}
\end{equation}
where $\Delta_{\Comp}=\Delta\setminus\{(\sigma_1,a,\sigma_2)\in\Sigma \mid a\in\Endg)\}$.
Any controller for \textbf{SG} would have to find and completely conduct a shortest plan for $(\sigma,\sigma')$ to reach $\sigma'$.
 
\section{Example: Fail-operational Driver Assistance}
\label{sec:example-bmw}

Elaborating on an example in \cite{Gleirscher2017a}, we apply our framework and algorithm to hazard analysis and elaboration of mitigation strategies.  We use the abbreviations introduced in \Cref{sec:backgr-motiv}.

\vspace{-1em}

\subsubsection{Identifying an Operational Situation.}
\label{sec:ident-oper-situ}

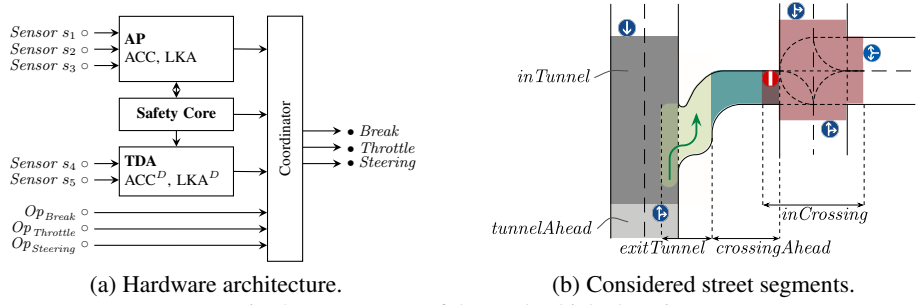
\begin{figure}[t]
  \subfloat[Hardware architecture.]{
{\footnotesize
\begin{tikzpicture}[scale=.65,every node/.style={transform shape},>=stealth,x=1.0cm, y=1.0cm]
  \coordinate (corig)   at (0,0);
  \coordinate (cS1) at (-1/2,0);
  \coordinate (cS2) at ($(-1/2,0)+(0,-1/3)$);
  \coordinate (cS3) at ($(-1/2,0)+(0,-2/3)$);
  
  \coordinate (cS4) at ($(-1/2,0)+(0,-8/3)$);
  \coordinate (cS5) at ($(-1/2,0)+(0,-9/3)$);
  \coordinate (cS6) at ($(-1/2,0)+(0,-11/3)$);
  \coordinate (cS7) at ($(-1/2,0)+(0,-12/3)$);
  \coordinate (cS8) at ($(-1/2,0)+(0,-13/3)$);
  \coordinate (cA1) at ($(-1/2,0)+(10/2,-6/3)$);
  \coordinate (cA2) at ($(-1/2,0)+(10/2,-7/3)$);
  \coordinate (cA3) at ($(-1/2,0)+(10/2,-8/3)$);
  
  \coordinate (cAP)   at (0,-3/3);
  \coordinate (cM)    at (0,-6/3);
  \coordinate (cADAS) at (0,-10/3);
  \coordinate (cCOORD) at (6/2,-14/3);

  \node[anchor=east] (S1) at (cS1) {$\mathit{Sensor}\ s_1\ \circ$};
  \node[anchor=east] (S2) at (cS2) {$\mathit{Sensor}\ s_2\ \circ$};
  \node[anchor=east] (S3) at (cS3) {$\mathit{Sensor}\ s_3\ \circ$};
  \node[draw, minimum width=2.1, minimum height=1.33cm, anchor=south west, text width=2.1cm, align=left] (AP) at (cAP) {\textbf{AP}\\ACC, LKA}; %
  \draw[->] (S1.east) -- node[above]{} ($(cAP) + (0, 3/3)$);
  \draw[->] (S2.east) -- node[above]{} ($(cAP) + (0, 2/3)$);
  \draw[->] (S3.east) -- node[above]{} ($(cAP) + (0, 1/3)$);
  \draw[->] (AP.east) -- node[above]{} ($(cCOORD) + (0,13/3)$);
  
  \node[draw, minimum width=2.1cm, minimum height=0.66cm, anchor=south west, text width=2.1cm, align=center] (M) at (cM) {\textbf{Safety Core}};
  \node[anchor=east] (S4) at (cS4) {$\mathit{Sensor}\ s_4\ \circ$};
  \node[anchor=east] (S5) at (cS5) {$\mathit{Sensor}\ s_5\ \circ$};
  \draw[->] (M.east) -- node[above]{} ($(cCOORD) + (0,9/3)$);
  
  \draw[<->] (AP.south) -- (M.north);
 
  \node[draw, minimum width=2.1cm, minimum height=1.0cm, anchor=south
  west, text width=2.1cm, align=left] (ADAS) at (cADAS) {\textbf{TDA}\\ACC$^D$, LKA$^D$};  
  \draw[->] (S4.east) -- node[above]{} ($(cADAS) + (0,2/3)$);
  \draw[->] (S5.east) -- node[above]{} ($(cADAS) + (0,1/3)$);
  \draw[->] (ADAS.east) -- node[above]{} ($(cCOORD) + (0,11/6)$);
  \draw[->] (M.south) -- (ADAS.north);
 
  \node[anchor=east] (S6) at (cS6) {$\mathit{Op}_\mathit{Break}\ \circ$};
  \node[anchor=east] (S7) at (cS7) {$\mathit{Op}_\mathit{Throttle}\ \circ$};
  \node[anchor=east] (S8) at (cS8) {$\mathit{Op}_\mathit{Steering}\ \circ$};
  \draw[->] (S6.east) -- node[above]{} ($(cCOORD) + (0,3/3)$);
  \draw[->] (S7.east) -- node[above]{} ($(cCOORD) + (0,2/3)$);
  \draw[->] (S8.east) -- node[above]{} ($(cCOORD) + (0,1/3)$);

  \node[draw,  minimum height=5.0cm, anchor=south west, text width=0.5cm, align=center] (COORD) at (cCOORD) {\rotatebox{90}{Coordinator}};

  \node[anchor=west] (A1) at (cA1) {$\bullet\ \mathit{Break}$};
  \node[anchor=west] (A2) at (cA2) {$\bullet\ \mathit{Throttle}$};
  \node[anchor=west] (A3) at (cA3) {$\bullet\ \mathit{Steering}$};
    
  \draw[->] ($(COORD.south east) + (0,8/3)$) -- node[above]{} (A1.west);
  \draw[->] ($(COORD.south east) + (0,7/3)$) -- node[above]{} (A2.west);
  \draw[->] ($(COORD.south east) + (0,6/3)$) -- node[above]{} (A3.west);
\end{tikzpicture}
}
     \label{fig:bna}}
  \hfill
  \subfloat[Considered street segments.]{
    \includegraphics[scale=.7]{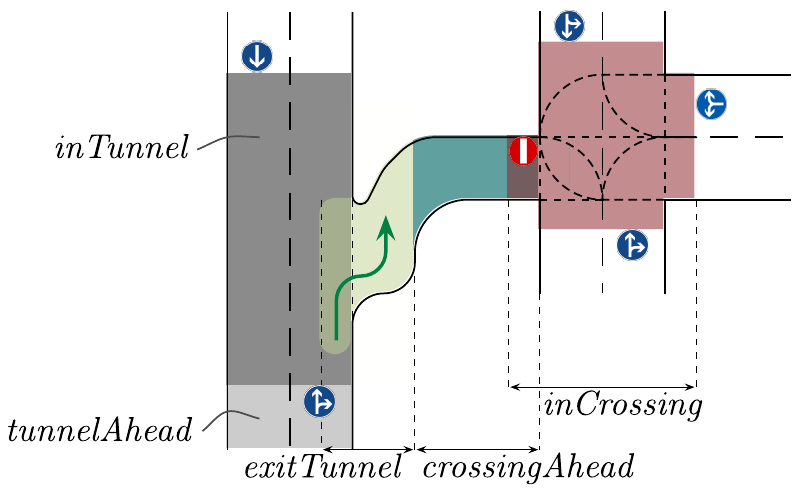}
    \label{fig:tunnel_crossing}}
  \caption{Two cutouts of the road vehicle domain.}
\end{figure}

We consider the situation $os\in\Os$: ``AV is taking an exit in a tunnel, at a speed between 30 and 90 km/h, with the driver being properly seated, and the next road segments contain a crossing.''  \Cref{fig:tunnel_crossing} depicts the corresponding street segment. 

\begin{table}[b]
  \centering
  \caption{Exemplary state variables of the different domains.}
  \label{tab:statevariables}
  \begin{tabularx}{1.0\linewidth}{lXl}
  \toprule\noalign{\smallskip}
  \textbf{Domain} & \textbf{State Variables} & \textbf{Abbreviation}\\
  \midrule\noalign{\smallskip}
  Driver  & Physical presence, consciousness, vigilance, $\ldots$ & \texttt{drv}\\
  Vehicle & Speed, loc(ation), fault conditions, $\ldots$ & \texttt{veh}\\
    RoadEnv & Daylight, weather, traffic, road, $\ldots$ & \texttt{renv}\\
  \bottomrule
\end{tabularx}
\end{table}

\subsubsection{Modeling the Road Vehicle Domain.}
\label{sec:model-road-vehicle}

\Cref{fig:bna} shows a simplified control system architecture used for driver assistance systems.  We model the relevant state information according to the abstractions described in~\Cref{sec:modelinggeneral}.  State variables commonly used for road vehicles are listed in~\Cref{tab:statevariables}.
For $\Sigma^s$, we assume to have the  variables\footnote{Variable types and usage depend on the AV sensors and car2X services through which they are measured. We assume individual error estimators for all variables.} (prefixed with their domains, in parentheses their types):
$\mathtt{veh.loc}$ (coordinate),
$\mathtt{veh.speedvec}$ (vector of floats), 
$\mathtt{renv.map}$ (street map\footnote{With, \egs
    topological coordinate system, information about tunneled parts.}), and
$\mathtt{drv.pos}$ (enumeration).  \texttt{veh} denotes all variables of this domain.  For $\Sigma^p$, we identify the following predicates%
\footnote{Here, $P_x$ refers to a pattern for the street map element class $x$ which acts like a filter on the street map data type.
For sake of brevity, we omit details of sensor fusion and street map calculations required for evaluating these predicates.}:
\begin{eqnarray*}
  \label{eq:1}
  \mathit{exitTunnel} &\equiv& 
    \mathtt{veh.route}\subset\mathtt{renv.map}\cap (P_{\mathrm{exit}} \cup P_{\mathrm{tunnel}})\\
  \mathit{crossingAhead} &\equiv& 
    \mathtt{veh.route}\cap(\mathtt{renv.map}\cap P_{\mathrm{crossing}})\neq\emptyset\\
  \mathit{drvSeated} &\equiv& \mathtt{drv.pos} = \mathrm{seated}
\end{eqnarray*}
Furthermore, we use unspecified predicates: 

\vspace{-1.5em}
\begin{align*}
  \mathit{inTunnel} &\equiv p_4(\mathtt{veh.loc}, \mathtt{renv.map}) &
  \mathit{A}        &\equiv p_5(\mathtt{veh.faults})\\
  \mathit{L}        &\equiv p_6(\mathtt{veh.faults}) &
  \mathit{R}        &\equiv p_7(\mathtt{drv.vigilance})\\
  \mathit{inCrossing} &\equiv p_x(\mathtt{veh.loc}, \mathtt{renv.map}) & 
  \mathit{tunnelAhead} &\equiv p_y(\mathtt{veh.loc}, \mathtt{renv.map})
\end{align*}

\noindent
The invariant for $\os$ is
\label{sec:spec-state-invar}
  $p_{\os} \equiv exitTunnel \land
               drvSeated \land
      crossingAhead$.
Note that the AP is active in the initial state $\sigma_0$ associated with $\os$. 

\paragraph{Notation.} In the following  (\Cref{fig:riskgraph-iter2,fig:riskgraph-iter2-simpl,fig:riskgraph-iter3}),
for each state, 
$H$ denotes that the hazard $H$ is active (phase $e^H$), $\underline{H}$ that $H$ contributed to a mishap (phase $e^H_m$, only in \Cref{tab:details-iter2}), 
and $\overline{H_i}$ that its $i$th mitigation phase is active (phase $m^H_i$). 
We do not indicate hazards which are in phase $0$.

\subsubsection{Incremental Forward Construction of the Risk Structure.}

Refining the regions $\sthaz$ and $\stsaf$~(\Cref{fig:statemodel}), we construct $\RS$ from three hazards $A,L$, and $R$ identified by $\mathsf{hazId}$~(\Cref{sec:hazid}).
\Cref{tab:details-iter2} sketches the construction of the first and second increments towards $\RS_2$, including the events $A\equiv$``AP sensor $s_1$ fault'' and $L\equiv$``TDA LKA$^D$ software fault.''

\begin{table}[t]
  \centering
  \caption{Model after two increments ($\RS_{2}$). $\tpar$ denotes true parallelism, $;$ concatenation.}
  \label{tab:details-iter2}
  \resizebox{.9\textwidth}{!}{
\begin{tabularx}{1.0\textwidth}{cX@{\hspace{1em}}X}
  \toprule
  \textbf{1+2} & 
  \textbf{Description} & 
  \textbf{Model Increment} 
  \\\midrule
  
  $\Sigma^s$ 
    & Introduce faults (\egs from fault model)
      & $\mathtt{veh.faults}$\\
  $\Haz$ & AP sensor $s_1$ fault 
      & $A \equiv p_5(\mathtt{veh.faults})$ \\
  $\Haz$ & TDA LKA$^D$ software fault
      & $L\equiv p_6(\mathtt{veh.faults})$\\
  $\Sigma$ & End.\ phases: Comb.\ of $A$ and $L$ 
      & $AL\equiv A \land L$, $\stc[1]{A}L \equiv \stc[1]{A} \land L$ \\
    & \multicolumn{2}{p{.9\linewidth}}{$\quad$\hfill$\stc[1]{A}L$ \dots ``LKA$^D$ faulty'' $\land$ ``TDA active'' $\land$ ``AP out of the loop''} 
  \\\midrule

  $\Endg$ 
    & Actions establishing $A$ and $L$ (\egs from
      architecture analysis) 
      & $\acf[A]$, $\acf[L]$ \hfill $\Endg=\{\acf[A],\acf[L]\}$ 
  \\\midrule

  $\Weights$ 
    & Probability of endangerment %
      & \egs $\mathsf{pr}(\acf[A]) := .01$, %
        $\mathsf{pr}(\acf[L]) := .02$ \\
    & Severity %
      \hfill $\stmh{AL}$\dots ``high-speed collision''
      & $\stmh{AL}\in\stmis$, $\sev{\stmh{AL}}=f$  %
  \\\midrule

  $\Comp$ 
    & $\accomp[1]{A}$\dots ``AP fail-op.\ by degrad.\ to TDA'' 
      & $\accomp[1]{A}\equiv\ac[TDA]{fo}$ \\
    & $\accomp[2]{A}$\dots ``deact.\ ACC'' $\tpar$ ``driver in loop'' 
      & $\accomp[2]{A}\equiv\ac[ACC^D]{off}\tpar\ac[Drv]{on}$ \\
    & $\accomp[3]{A}$\dots ``AP fail-silent'' 
      & $\accomp[3]{A}\equiv\ac[AP]{fs}\tpar\ac[Drv]{on}$ \\
    & $\accomp[1]{L}$\dots ``TDA fail-silent and warn'' 
      & $\accomp[1]{L}\equiv\ac[L2]{fs};\ac[L2]{warn}$ \\
    & $\accomp[2]{L}$\dots ``TDA total fail-silent'' $\tpar$ ``immediate handover to driver'' 
      & $\accomp[2]{L}\equiv\ac[*]{fs}\tpar\ac[Drv]{on}$,
        $\accomp[3]{L} \equiv \accomp[4]{L} \equiv \accomp[2]{L}$ 
  \\\midrule

  $\Sigma$ & \multicolumn{2}{p{.85\linewidth}}{
    Mitigation phases: $\stc[1]{A}$ \dots ``$s_1$ fault'' $\land$ ``TDA active,'' \newline
    $\stc[2]{A}$ \dots ``$s_1$ fault'' $\land$ ``handed to driver'' $\land$ ``TDA active'' \newline
    $\stc[3]{A}$ \dots ``$s_1$ fault'' $\land$ ``handed to driver'' $\land$ ``AP out of the loop'' \newline
    $\stc[1]{L}$ \dots ``TDA out of the loop'' $\land$ ``driver warned'' \newline
    $\stc[1]{A}\stc[1]{L}$ \dots ``AP and TDA out of the loop'' $\land$ ``handed to driver'' 
  }\\\midrule

  $\Weights$ 
    & Probability of mitigation 
      & \egs $\mathsf{pr}(\accomp[3]{A}) = .50$ \\
    & Cost of mitigation 
      & \egs $\mathsf{cs}(\accomp[3]{A}) = 3$ 
  \\\midrule

  $\mathfrak{R}$  
    & \multicolumn{2}{l}{Simplifications: \egs
      $\stc[2]{A} \approx_m \stc[3]{A}$ (\cf\Cref{def:comp-equiv})}
  \\\bottomrule
\end{tabularx}

   }
\end{table}

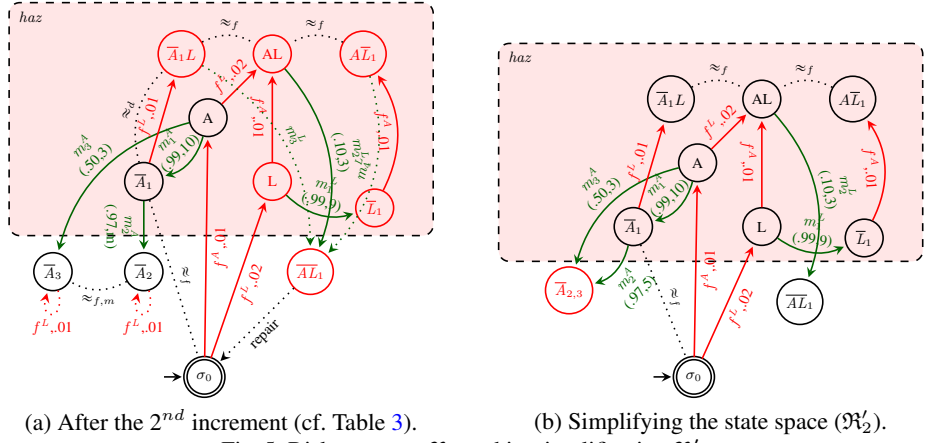
\begin{figure}[t]
  \subfloat[After the 2$^{nd}$ increment (cf.~\Cref{tab:details-iter2}).]{
\begin{tikzpicture}[
  scale=.63,every node/.style={transform shape},
  ->,>=stealth,semithick,node distance=6em,
  accomp/.style={sloped,DarkGreen,align=center},
  acendg/.style={sloped,red,align=center},
  equiv/.style={sloped,dotted},
  shorten >=1pt,auto,initial text={}]

  \node[draw,fill=red!10,rounded corners,minimum
  width=28em,minimum height=15.5em,dashed] (haz) at (1,15.5) {};
  \node at ($(haz.north west)+(1.5em,-1em)$) {$\sthaz$};

  \node[state,initial,accepting] (saf) {$\sigma_0$}; 
  \node[state] (A2) at (-4em,7em) {$\stc[2]{A}$}; 
  \node[state] (A1) [above of=A2] {$\stc[1]{A}$}; 
  \node[state] (A3) [left of=A2] {$\stc[3]{A}$}; 
  \node[state] (A) [above right of=A1] {A}; 

  \node[state,red] (LA) [above right of=A] {AL}; 
  \node[state,red] (L1A) [right of=LA] {$A\stc[1]{L}$}; 
  \node[state,red] (LAc1) [left of=LA] {$\stc[1]{A}L$}; 
  \node[state,red] (L) [below right of=A] {L}; 
  \node[state,red] (LA1) at (7em,7em) {$\stc[1]{AL}$}; 
  \node[state,red] (L1) [above right of=LA1] {$\stc[1]{L}$}; 

  \path[acendg] 
  (saf) edge node[below] {$\acf[A]$,.01}      (A);
  \path[accomp]
  (A) edge[bend left] node[above] {$\accomp[1]{A}$\\ (.99,10)} (A1)
  (A1) edge node[below] {$\accomp[2]{A}$\\ (.97,m)} (A2)
  (A) edge[bend right] node[above] {$\accomp[3]{A}$\\ (.50,3)} (A3);
  \path[equiv] 
  (A1) edge[-] node[above] {$\approx_f$} (saf);

  \path[acendg] 
  (saf) edge node[below] {$\acf[L]$,.02}      (L)
  (L) edge node[below] {$\acf[A]$,.01}        (LA)
  (A2) edge[dotted,loop below] node[below] {$\acf[L]$,.01}        (A2) %
  (A3) edge[dotted,loop below] node[below] {$\acf[L]$,.01}        (A3)
  (A1) edge node[above] {$\acf[L]$,.01}        (LAc1)
  (L1) edge[bend right] node[below] {$\acf[A]$,.01}        (L1A)
  (A)   edge node[above] {$\acf[L]$,.02}      (LA);
  \path[accomp]
  (L) edge[bend right] node[above] {$\accomp[1]{L}$\\ (.99,9)} (L1)
  (LAc1) edge[dotted,bend left] node[above] {$\accomp[3]{L}$} (LA1)
  (LA) edge[bend left] node[above] {$\accomp[2]{L}$\\ (.10,3)} (LA1)
  (L1A) edge[dotted,bend left] node[above] {$\accomp[4]{L}$} (LA1);
  \path[equiv] 
  (A1) edge[-,bend left] node[above] {$\approx_d$} (LAc1)
  (L1A) edge[-,bend right] node[above] {$\approx_f$} (LA)
  (LAc1) edge[-,bend left] node[above] {$\approx_f$} (LA)
  (A2) edge[-,bend left] node[below] {$\approx_{f,m}$} (A3);
  \path
  (LA1) edge[dotted,sloped] node[below] {repair} (saf);
\end{tikzpicture}
     \label{fig:riskgraph-iter2}}
  \hfill
  \subfloat[Simplifying the state space ($\RS'_2$).]{
\begin{tikzpicture}[
  scale=.63,every node/.style={transform shape},
  ->,>=stealth,semithick,node distance=6em,
  accomp/.style={sloped,DarkGreen,align=center},
  acendg/.style={sloped,red,align=center},
  equiv/.style={sloped,dotted},
  shorten >=1pt,auto,initial text={}]

  \node[draw,fill=red!10,rounded corners,minimum
  width=28em,minimum height=14.5em,dashed] (haz) at (1,13.5) {};
  \node at ($(haz.north west)+(1.5em,-1em)$) {$\sthaz$};

  \node[state,initial,accepting] (saf) {$\sigma_0$}; 
  \node[state] (A1) at (-4em,10em) {$\stc[1]{A}$}; 
  \node[state,red] (A2) [below left of=A1] {$\stc[2,3]{A}$}; 
  \node[state] (A) [above right of=A1] {A}; 

  \node[state] (LA) [above right of=A] {AL}; 
  \node[state] (L1A) [right of=LA] {$A\stc[1]{L}$}; 
  \node[state] (LAc1) [left of=LA] {$\stc[1]{A}L$}; 
  \node[state] (L) [below right of=A] {L}; 
  \node[state] (LA1) at (7em,5em) {$\stc[1]{AL}$}; 
  \node[state] (L1) [above right of=LA1] {$\stc[1]{L}$}; 

  \path[acendg] 
  (saf) edge node[below] {$\acf[A]$,.01}      (A);
  \path[accomp]
  (A) edge[bend left] node[above] {$\accomp[1]{A}$\\ (.99,10)} (A1)
  (A1) edge[bend left] node[below] {$\accomp[2]{A}$\\ (.97,5)} (A2)
  (A) edge[bend right] node[above] {$\accomp[3]{A}$\\ (.50,3)} (A2);
  \path[equiv] 
  (A1) edge[-] node[above] {$\approx_f$} (saf);

  \path[acendg] 
  (saf) edge node[below] {$\acf[L]$,.02}      (L)
  (L) edge node[below] {$\acf[A]$,.01}        (LA)
  (A1) edge node[above] {$\acf[L]$,.01}        (LAc1)
  (L1) edge[bend right] node[below] {$\acf[A]$,.01}        (L1A)
  (A)   edge node[above] {$\acf[L]$,.02}      (LA);
  \path[accomp]
  (L) edge[bend right] node[above] {$\accomp[1]{L}$\\ (.99,9)} (L1)
  (LA) edge[bend left] node[above] {$\accomp[2]{L}$\\ (.10,3)} (LA1);
  \path[equiv] 
  (LAc1) edge[-,bend left] node[above] {$\approx_f$} (LA)
  (L1A) edge[-,bend right] node[above] {$\approx_f$} (LA);

\end{tikzpicture}
     \label{fig:riskgraph-iter2-simpl}}
  \caption{Risk structure $\RS_2$ and its simplification $\RS'_2$.}
\end{figure}

\Cref{fig:riskgraph-iter2} shows $\Delta$ for $\RS_2$.  According to \Cref{alg:consriskgraph}, we try to add the fault condition $L$ to $\sigma_0$ and other states in $\RS_1$ (\ies black states in \Cref{fig:riskgraph-iter2}).  Based on the action $\acf[L]$, this step yields the states $L,\stc[1]{A}L$, and $AL$. Then, a mitigation step yields the states $\stc[1]{L}$ and $\stc[1]{AL}$ and, finally, another step of endangerment analysis based on the action $\acf[A]$ yields $A\stc[1]{L}$.

\paragraph{Risk Priority Estimation.}
From the state $\stmh{AL}$ with $\sev{\stmh{AL}}=f$, we can derive, \egs $\riskprio{\stc[1]{A}}$ according to Eq.~\eqref{eq:riskprio}.  We can as well derive $\riskprio{\stc[2]{A}} = \riskprio{\stc[3]{A}} = m$ because reaching $\stmh{AL}$ by driving assistance control is no more possible.

\paragraph{Equivalences and Model Reduction.}
\label{sec:simpl-comp-state}
In \Cref{fig:riskgraph-iter2}, \eg
\begin{itemize}
\item $\stc[2]{A} \approx_m \stc[3]{A}$ because in both states $A$ is mitigated and other hazards are inactive ($0$, 
\cf\Cref{def:comp-equiv}),
\item $\stc[1]{A} \approx_f \sigma_0$ because in $\stc[1]{A}$ the degraded variants of LKA and ACC, \ies LKA$^D$ and ACC$^D$, are in the loop,
\item $\stc[1]{A} \approx_d \stc[1]{A}L$ because in both states LKA$^D$ and ACC$^D$ are in the loop, 
\item $\stc[1]{A}L \approx_f AL$ because in both states, LKA and ACC are in the loop, and
\item $\stc[1]{A}L \not\approx_h AL$ because ACC (part of AP) is faulty and ACC$^D$ (part of TDA) is fully operational.
\end{itemize}
Simplifications can be derived from \Cref{fig:riskgraph-iter2}, where we might (i) merge two states $(\sigma_1,\sigma_2)\in\;\approx_d$ if $\riskprio{\sigma_1}=\riskprio{\sigma_2}$, or (ii) merge two consecutive states on a ``safe'' mitigation path, \egs from any $\sigma\in\sthaz$ to $\sigma_0$ if actions such as \emph{limp-home}, \emph{shutdown}, and \emph{repair} are feasible from $\sigma$.

\Cref{fig:riskgraph-iter2-simpl} shows a simplification $\RS'_2$ of $\RS_2$.  We omit irrelevant transitions ($\acf[L]$) and collapse the mitigation-equivalent ($\approx_m$) states $\stc[2]{A}$ and $\stc[3]{A}$.  Consequently, with the states $\stc[2,3]{A}$ and $\stc[1]{AL}$ we get a refinement of $\stsaf$.
According to~Eq.~\eqref{eq:safpassstat}, $\stc[2]{A}$ is a \emph{safest possible state} reachable from $A$.

\begin{figure}[t]
  \centering
\begin{tikzpicture}[
  scale=.6,every node/.style={transform shape},
  ->,>=stealth,semithick,node distance=6em,
  accomp/.style={DarkGreen},
  acendg/.style={red},
  equiv/.style={sloped,dotted},
  shorten >=1pt,auto,initial text={}]

  \node[state,initial,accepting] (saf) {$\sigma_0$}; 
  \node[state] (A) [above of=saf] {A}; 

  \node[state] (AL) [above of=A] {AL}; 
  \node[state] (L) [left of=A] {L}; 

  \node[state,red] (R) [right of=A] {$R$}; 
  \node[state,red] (LR) [above of=L] {$LR$}; 
  \node[state,red] (AR) [above of=R] {$AR$}; 
  \node[state,red] (ALR) [above of=AL] {$ALR$}; 

  \node[state] (A1) [left of=L] {$\stc[1]{A}$}; 
  \node[state] (A2) [left of=A1] {$\stc[2,3]{A}$}; 
  \node[state] (AL1) [right of=AR] {$A\stc[1]{L}$}; 
  \node[state] (A1L) [left of=LR] {$\stc[1]{A}L$}; 
  \node[state] (L1) [right of=R] {$\stc[1]{L}$}; 

  \node[state,red] (A1R) [above of=A2] {$\stc[1]{A}R$}; 
  \node[state,red] (A1LR) [above of=A1L] {$\stc[1]{A}LR$}; 
  \node[state,red] (AL1R) [above of=AL1] {$A\stc[1]{L}R$}; 
  \node[state] (A1L1) [below right of=AL1R] {$\stc[1]{A}\stc[1]{L}$}; 
  \node[state,red] (L1R) [above right of=L1] {$\stc[1]{L}R$}; 

  \node[state,DarkGreen] (A1L1R) [left of=A1LR] {$\stc[1]{A}\stc[1]LR$}; 

  \path[acendg,black] 
  (saf) edge node[below left] {$\acf[A]$}      (A);
  \path[accomp,black]
  (A) edge[bend left] node[below] {$\accomp[1]{A}$} (A1)
  (A1) edge node[below] {$\accomp[2]{A}$} (A2)
  (A) edge[bend right] node[above] {$\accomp[3]{A}$} (A2);
  \path[equiv] 
  (A1) edge[-,bend right] node[below] {$\approx_f$} (saf);

  \path[acendg,black] 
  (saf) edge node {$\acf[L]$}      (L)
  (L) edge node {$\acf[A]$}        (AL)
  (A1) edge node[above left] {$\acf[L]$}        (A1L)
  (L1) edge node {$\acf[A]$}        (AL1)
  (A)   edge node[below left] {$\acf[L]$}      (AL);
  \path[accomp,black]
  (L) edge[bend right] node[below right] {$\accomp[1]{L}$} (L1)
  (AL) edge[bend left] node[above] {$\accomp[2]{L}$} (A1L1);
  \path[equiv] 
  (A1L) edge[-,bend left] node[above] {$\approx_f$} (AL)
  (AL1) edge[-,bend right] node[above,near start] {$\approx_f$} (AL);

  \path[acendg] 
  (saf) edge node {$\acendg{R}$} (R)
  (L) edge node[below left] {$\acendg{R}$} (LR)
  (A) edge node[below] {$\acendg{R}$} (AR)
  (AL) edge node {$\acendg{R}$} (ALR)
  (A1) edge node {$\acendg{R}$} (A1R)
  (A1L) edge node {$\acendg{R}$} (A1LR)
  (AL1) edge node {$\acendg{R}$} (AL1R)
  (L1) edge[bend right] node {$\acendg{R}$} (L1R)

  (R) edge node[right] {$\acf[A]$} (AR)
  (R) edge node[above right] {$\acf[L]$} (LR)
  (AR) edge node {$\acf[L]$} (ALR)
  (LR) edge node {$\acf[A]$} (ALR)
  ;

  \path[accomp,DarkGreen]
  (A1LR) edge[bend right] node[above] {$\accomp[3]{L}$} (A1L1R);
\end{tikzpicture}
   \caption{Risk structure after adding endangerments (in red) for the 3$^{rd}$ increment (weights not shown, cf.~\Cref{tab:details-iter3}).}
  \label{fig:riskgraph-iter3}
\end{figure}
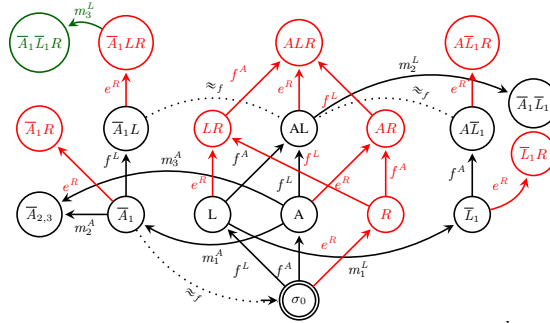

Next, \Cref{tab:details-iter3} and \Cref{fig:riskgraph-iter3} describe a cut-out of $\RS_3$ after the third increment where we added the event $R\equiv$``Driver reaction time increases.''  

\begin{table}[t]
  \centering
  \caption{Adding endangerments for the third increment ($\RS_3$).}
  \label{tab:details-iter3}
  \resizebox{.9\textwidth}{!}{
  \begin{tabularx}{1.0\linewidth}{cX@{\hspace{1em}}X}
    \toprule
    \textbf{3} & \textbf{Description} & \textbf{Model Increment} 
    \\\midrule
    $\Haz$  
      & Driver reaction time increases. 
      & $R \equiv p_7(\mathtt{drv.vigilance})$  
    \\\midrule
    $\Sigma$ & States & 
      $R, LR, AR, ALR$, \\
        & & $\stc[1]{A}R$,
            $\stc[1]{A}LR$,
            $A\stc[1]{L}R$,
            $\stc[1]{L}R$
    \\\midrule
    $\Endg$   & 
    Action $\acendg{R}$ \dots
          ``driver looks sidewards''\newline $\tpar$ ``hands go off steering wheel''
        & $\acendg{R}$ \hfill 
        $\Endg = \{\acf[A],\acf[L],\acendg{R}\}$
    \\\midrule
    $\Comp$ & &
    $\accomp[3]{L} \equiv warn \parallel_t normalStop$ 
    \\\bottomrule
  \end{tabularx}}
\end{table}

\section{Discussion of Limitations, Applicability, and Strengths}
\label{sec:discussion}

The abstraction $\sstates$~(\Cref{sec:abstractions}) is subject to standard signal processing steps, \ies \emph{sampling} of continuous signals at discrete time points,
\emph{quantization} of dense domains to form finite domains, and
\emph{clamping} of domains. %
We assume all signals to be sampled faster then their respective \textsc{Nyquist} period, sufficiently small quantums, and sufficiently large ranges of data types.  Furthermore, we expect a mitigation planner to be fast enough (sufficiently low latency) to provide outputs for effective and optimal control.
Note that the risk structure abstracts from the low-level parameters necessary for actual control of mitigations which takes place at the level of $\sstates$.

The treatment of these issues will determine how 
accurate mitigations can take place at the right time and duration.
In addition, we might consider \emph{higher-order mitigations} to handle adverse impacts of first-order mitigations.  However, such impacts have to be identified as hazards to get recognized in $\RS$.

Elaborating on risk regions~(\Cref{sec:regions}), $\stmis$ represents mitigation-less harmful states, however, $\sthaz$ includes all states where mitigations are feasible.  Consequently, we allow ``bad things to happen'' as long as we have \emph{partial mitigations}, \egs an airbag would prevent from reaching $\stmis$ at a certain probability.

\section{Conclusion and Future Work}
\label{sec:conclusion}

We presented \emph{risk structures} as a model to design high-level controllers capable of run-time hazard mitigation, \ies of maintaining or reaching the safest states in a given operational situation.  We sketched an incremental approach to develop mitigation strategies.  Safety measures are a combination of reducing or eliminating endangerments with constructing or strengthening mitigations.  Risk structures can help to derive safety requirements for a control system architecture.  Moreover, they can lay a basis for the evaluation, choice, and combination of mitigation strategies.  Our example highlights challenges to tackle in hazard mitigation of fail-operational automated driving.  Finally, we indicate how several formalisms---temporal specification, predicate abstraction, and transition systems---can coherently aid in hazard mitigation planning.

\paragraph{Future Work.}
\label{sec:object-our-meth}

Based on risk structures, we aim to \emph{evaluate} criteria such as
\begin{inparaenum}[(i)]
\item time, energy, and cost of mitigations,
\item the role of human intervention,
\item resilience to change of operational situations,
\item control system simplicity.
\end{inparaenum}

In the next steps, we want to efficiently \emph{automate} the derivation of acceptable mitigation strategies, and synthesize feasible and affordable mitigation strategies.
Based on weights, we can define desirable properties of mitigation strategies implemented in $\RS$, \egs monotonicity.

\begin{definition}[Mitigation Monotonicity]
  Let $S\subset\Delta^{\infty}$ be a strategy (\Cref{def:strategy}) and $n\in\mathbb{N}\setminus\{0\}$.  We call $S$ \emph{mitigation monotonous} iff for each path $\sigma_0\stackrel{a_0}{\longrightarrow}\ldots\stackrel{a_{n-1}}{\longrightarrow}\sigma_n \in S \colon \forall i\in\{0,\ldots,n-1\} \colon \riskprio{\sigma_i} \geq_{\wsev} \riskprio{\sigma_{i+1}}$.
\end{definition}
Intuitively, during planning we seek mitigation paths containing only endangerments, if any, which do not increase risk priority.  This might, however, be a definition to be relaxed for practical use by, \egs allowing $\wriskprio$-distances.

Given that we \emph{use our algorithm off-line}, it is important to 
 make the $\mathsf{poss}$ and $\procdet$ steps in \Cref{alg:consriskgraph} interactive for the safety engineer.
Moreover, instead of elaborating $\os$-specific risk structures off-line, we aim at using our algorithm to generate such structures on-line given a specific operational situation, and combine this with a transition system switching between operational situations.
Given that we \emph{use our algorithm on-line}, it is important to 
develop simplification rules to be applied to $\Sigma$ based on the equivalences in \Cref{sec:concepts-ocp}.

We plan to evaluate our results in the automotive industry whose aims include checking whether fail-operational extensions of given in-vehicle network architectures for automated driving can be made acceptably safe.

Finally, for a \emph{regulatory agency} to apply our approach to AV, we have to show (i) our approach using a large example involving several operational situations, (ii) how our abstraction can be verified, and (iii) that the limits of controllers do not constrain our approach to achieve safe stable control loops.

\vspace{-1em}

\small
\subsubsection*{Acknowledgments.}
We are grateful to Maximilian Junker for a thorough review of this
work.  Moreover, we thank our project partners from the German
automotive industry for inspiring discussions and providing a highly
innovative practical context for our research.  Furthermore, we thank
our peer reviewers for suggestions on the use of risk structures,
signal processing, and regulatory certification.

\vspace{-1em}

\bibliography{../literature}

\end{document}